# Asynchronous Interference Alignment

Mehdi Torbatian, Hossein Najafi, and Mohamed Oussama Damen

Department of Electrical and Computer Engineering

University of Waterloo

Waterloo, ON, Canada, N2L 3G1

Email: {m2torbat, hnajafi, mdamen}@uwaterloo.ca

## Abstract

A constant $K$-user interference channel in which the users are not symbol-synchronous is considered. It is shown that the asynchronism among the users facilitates aligning interfering signals at each receiver node while it does not affect the total number of degrees of freedom (DoF) of the channel. To achieve the total $K/2$ DoF of the channel when single antenna nodes are used, a novel practical interference alignment scheme is proposed wherein the alignment task is performed with the help of asynchronous delays which inherently exist among the received signals at each receiver node. When each node is equipped with $M > 1$ antennas, it is argued that the same alignment scheme is sufficient to achieve the total $MK/2$ DoF of the medium when all links between collocated antennas experience the same asynchronous delay.

## I. INTRODUCTION

Asynchronism inherently exists in many communication systems mainly due to the effect of multi-path and propagation delay. As a fundamental issue in design of communication systems, the asynchronism can detrimentally affect the system performance if it is ignored or not dealt with properly. Although in most cases, it is presumed that terminals are synchronized by an infrastructure service provider, this assumption cannot be held in many communication scenarios such as the interference channel in which multiple independent receivers are randomly distributed over a geometrical area. In such scenarios, only one of the receivers can possibly receive a synchronized version of the transmitted signals (if an infrastructure synchronizing center exists) and the rest receive random asynchronous combinations of them. Contrary to intuition, some exceptions have been reported wherein the asynchronism has helped to improve the system performance [1]–[3]. For example, it is shown in [3] that in contrast to the synchronous pulse amplitude modulation (PAM), asynchronous PAM can exploit the total existing degrees of freedom of a multiple-input multiple output (MIMO) channel which communicates over a spectral mask with infinite support. In this work, we investigate the effect of the asynchronism naturally existing among the users on the total number of DoF of the $K$-user symbol-asynchronous interference channel.

### A. Prior Works

Despite many efforts during the last thirty years [4]–[8], the capacity region of the Gaussian interference channel is not known yet. To approximate the capacity at least for large values of signal to noise ratio (SNR, $\rho$), the region of DoF and the total number of DoF of the channel are studied. The DoF region, which roughly determines the shape of the capacity region at high values of SNR, is defined as the ratio of the capacity region and $\log \rho$ when $\rho \to \infty$. Similarly, the total DoF is defined as the ratio of the sum-capacity and $\log \rho$ when $\rho \to \infty$. In [9], it is shown that the total DoF of the $K$-user synchronous interference channel is $K/2$ when fading coefficients are time-varying, that is, each pair can benefit from half of its original degree of freedom with no interference from other pairs. This upper-bound is achieved by a technique called *Interference Alignment*. The key idea of interference alignment is to design signals such that they overlap at non-intended receivers while remaining distinct from the interferences at desired receivers [10]. It is shown in [11] that in ergodic scenarios, a simple paring scheme of particular channel matrices is adequate to achieve the total $K/2$ DoF of the channel.

In practice, however, the transmission rate is usually much faster than the rate of the channel variation resulting in quasi-static (constant) or block fading channel models. For such channels where the link coefficients are fixed for a long period of time, it is shown in [12] that the total multiplexing gain is upper bounded by $K/2$. Inspired by the idea of interference alignment for varying fading channels, many efforts have been made to generalize the idea to constant interference channels. In [13], it is shown that in a quasi-static $K$-user interference channel with real fading coefficients, there are $K/2$ degrees of freedom if the direct fading coefficients are irrational and the crossing gains are rational numbers, while it is strictly less than $K/2$ when all the coefficients are rational numbers. In [14], using asymmetric complex signaling over constant interference channel, it is shown that a minimum total DoF equal to 1.2 is achievable for almost all values of channel coefficients (outside a subset of measure zero) of an interference channel with three or more users. Inspired by [13], the authors in [15] take advantage of the properties of the rational and the irrational numbers to align the users' interferences in the constant $K$-user interference channel to achieve the total $K/2$ DoF of this channel. While the proposed scheme, which is known as real-interference-alignment, performs the alignment task almost surely at infinite SNR with infinite quantization precision, it does not offer a feasible solution for practical ranges of SNR.

A common unrealistic assumption in all of the aforementioned schemes is that the users are fully synchronous and the received signal at each receiver node is a synchronized linear combination of the transmitted signals. In [16] interference alignment based on propagation delays of signals is proposed via an example of proper node placement in a wireless network containing four nodes. In [17], a fully connected $K$-user interference channel is modeled by a time indexed interference graph and the alignment task is associated with finding the maximal independent set of the graph via a dynamic programming algorithm. As a simplifying assumption, asynchronous delays among the received signals at each receiver are assumed to be integer multiples of the symbol interval.

### B. System Description

In this paper[1], we consider a symbol-asynchronous $K$-user interference channel and investigate the effect of the asynchronism on the total DoF of this channel. It is argued that the asynchronism facilitates aligning interfering signals at each receiver node while it does not affect the total DoF of the channel. Each node is equipped with a single antenna. The channel coefficient between the $j$-th transmitter and the $i$-th receiver is denoted by $h_{i,j}$, $i,j \in \{1,\ldots,K\}$. We assume that the fading coefficients are constant non-zero random variables drawn from a probability distribution. The nodes are assumed to be frame-synchronous, however, not symbol-synchronous, i.e., the beginning and the end of each frame align up to a delay of length less than a symbol interval, $T_s$. Let $\tau_{i,j}$ denote the relative delay of the received signal from the $j$-th transmitter to the $i$-th receiver. We assume $\tau_{i,j}$'s are i.i.d random variables drawn from a continuous probability distribution and $0 \leq \tau_{i,j} < T_s$, $\forall i,j \in \{1,2,\ldots,K\}$. They are constant during

---

[1]This work has partly been published in [18].

transmission of a frame. Let

$$\tau_{m,j}^{[i]} \triangleq \tau_{i,m} - \tau_{i,j}, \quad \forall i,j,m \in \{1,2,\ldots,K\}, \qquad (1)$$

be the relative delay between the transmitted signals from the $m$-th and the $j$-th transmitters measured at the $i$-th receiver node. Since $\tau_{i,j}$'s are continuous independent random variables over a symbol interval, $\tau_{m,j}^{[i]}$'s are distinct with probability one $\forall i,j,m \in \{1,2,\ldots,K\} (m \neq j)$ and $-T_s < \tau_{m,j}^{[i]} < T_s$.

Each transmitter uses a unit energy shaping waveform $\psi_j(t)$, $j \in \{1,\ldots,K\}$, to linearly modulate its information symbols in a PAM like signal. The transmitted signal of the $j$-th transmitter is given by

$$x_j(t) = \sum_k x_j(k)\psi_j(t - kT_s), \qquad (2)$$

where $x_j(k)$ is the transmitted symbol by the $j$-th transmitter at the $k$-th symbol interval. The received signal at the $i$-th receiver is given by [1]

$$y_i(t) = \sum_{j=1}^{K} h_{i,j} x_j(t - \tau_{i,j}) + z_i(t), \ i \in \{1,\ldots,K\}, \qquad (3)$$

where $z_i(t)$ is an additive white Gaussian noise process with zero mean and power spectral density $\sigma^2$.

*C. Main Result*

*Theorem 1:* The total number of DoF of the underlying constant $K$-user symbol-asynchronous interference channel with single antenna nodes is upper bounded by $K/2$. This upper-bound is achieved for almost all values of the fading coefficients and propagation delays.

We first argue that the total DoF of the underlying channel is upper bounded by $K/2$. Then, a novel interference alignment scheme, which deploys the asynchronous delays among the users, is proposed to achieve the total $K/2$ DoF of the channel almost surely. The asynchronism causes inter-symbol-interference (ISI) among transmitted symbols by different transmitters. Therefore, the underlying quasi-static links are converted to ISI and accordingly to time-varying channels providing the required channel variation for the interference alignment. Our scheme is similar to the vector alignment scheme invented in [9] for varying fading channels; however, it is proposed for the constant symbol-asynchronous interference channels. When each node is equipped with $M$ antennas, we argue that the same scheme achieves the total $MK/2$ DoF of the medium provided that all links between collocated antennas experience the same asynchronous delay. This results in performing the alignment task in a smaller number of channel uses.

The rest of the paper is organized as follows. In Sections II, the system model and the signaling schemes are discussed. The proof of Theorem 1 is given in Section III. The asynchronous interference channel with multiple antenna nodes is considered in Section IV. The paper is concluded in Section V.

**Notations:** In this work, letters with underline $\underline{x}, \underline{X}$ denote vectors, and boldface uppercase letters $\mathbf{X}$ denote matrices. The superscripts $[\cdot]^\intercal, [\cdot]^\dagger$ denote the transpose and the conjugate transpose of the corresponding vector or matrix, respectively. $\mathbf{I}_n$ is the identity matrix of dimension $n$.

II. SYSTEM MODEL AND SIGNALING SCHEME

We assume that all shaping waveforms are the same, i.e., $\psi_j(t) = \psi(t)$, $\forall j$. In theory, $\psi(t)$ is a band-limited waveform with bandwidth $W$, e.g., the sinc waveform, $\texttt{sinc}(x) = \frac{\sin \pi x}{\pi x}$. In practice, however, using band-limited waveforms with infinite time support is not feasible. Hence, a truncated version of common waveforms is used. We consider both cases of using band-limited and time-limited waveforms and present a unified channel model for both cases.

## A. When $\psi(t)$ Is a Band-Limited Waveform

In this case, for codewords of length $N$, the received signal at the $i$-th receiver sampled at $t = kT_s + \tau_{i,i}$, $k = 0, 1, \ldots, N-1$, is given by

$$y_i(k) = \sum_{j=1}^{K} h_{i,j} \sum_{q=0}^{N-1} \gamma_{i,j}(k-q) x_j(q) + z_i(k), \tag{4}$$

where $\gamma_{i,j}(k) \triangleq \psi\left(kT_s + \tau_{i,j}^{[i]}\right)$, $z_i(k)$ is the sample of the noise at $t = kT_s + \tau_{i,i}$. The received samples can be written in a vector form as follows,

$$\underline{y}_i = \sum_{j=1}^{K} h_{i,j} \hat{\mathbf{\Gamma}}_{i,j} \underline{x}_j + \underline{z}_i, \tag{5}$$

where $\underline{x}_j = [x_j(0), x_j(1), \ldots, x_j(N-1)]^\mathsf{T}$, $\underline{y}_i = [y_i(0), y_i(1), \ldots, y_i(N-1)]^\mathsf{T}$, $\underline{z}_i = [z_i(0), z_i(1), \ldots, z_i(N-1)]^\mathsf{T}$, and

$$\hat{\mathbf{\Gamma}}_{i,j} = \begin{bmatrix} \gamma_{i,j}(0) & \gamma_{i,j}(-1) & \gamma_{i,j}(-2) & \cdots & \gamma_{i,j}(-N+1) \\ \gamma_{i,j}(1) & \gamma_{i,j}(0) & \gamma_{i,j}(-1) & \cdots & \gamma_{i,j}(-N+2) \\ \gamma_{i,j}(2) & \gamma_{i,j}(1) & \gamma_{i,j}(0) & \cdots & \gamma_{i,j}(-N+3) \\ \vdots & \vdots & \vdots & \ddots & \vdots \\ \gamma_{i,j}(N-1) & \gamma_{i,j}(N-2) & \gamma_{i,j}(N-3) & \cdots & \gamma_{i,j}(0) \end{bmatrix}. \tag{6}$$

As can be seen, the asynchronism among the users causes ISI among the transmitted symbols by different users which is represented by matrix $\hat{\mathbf{\Gamma}}_{i,j}$. We approximate this matrix with its asymptotically equivalent circulant matrix and show that the approximation error is negligible for large codeword length.

*Definition 1:* Two matrix sequences $\{\mathbf{A}_N\}$ and $\{\mathbf{B}_N\}$, $N = 1, 2, \ldots$, are said to be asymptotically equivalent, and denoted by $\{\mathbf{A}_N\} \sim \{\mathbf{B}_N\}$, if the following conditions are satisfied [19]:

1) $\exists Q < \infty$ such that $\forall N$, $\|\mathbf{A}_N\| < Q$ and $\|\mathbf{B}_N\| < Q$,
2) $\lim_{N \to \infty} |\mathbf{A}_N - \mathbf{B}_N| = 0$,

where $\|\mathbf{A}\|$ and $|\mathbf{A}|$ denote the strong norm and the weak norm of $\mathbf{A}$ given by [20]

$$\|\mathbf{A}\| = \max_{\underline{x}} \left[ (\underline{x}^\dagger \mathbf{A}^\dagger \mathbf{A} \underline{x}) / (\underline{x}^\dagger \underline{x}) \right]^{1/2},$$

$$|\mathbf{A}| = \left[ N^{-1} \mathtt{trace}(\mathbf{A}^\dagger \mathbf{A}) \right]^{1/2}.$$

It is shown in [19] that for all absolutely summable infinite complex sequences $\{\gamma_{i,j}(k), k \in \mathbb{Z}\}$ with $2\pi$-periodic discrete time Fourier transform (DTFT), $\Gamma_{i,j}(\omega)$, given as

$$\Gamma_{i,j}(\omega) = \sum_{k} \gamma_{i,j}(k) e^{-\xi \omega k}, \tag{7}$$

with $\xi = \sqrt{-1}$, $\{(\hat{\mathbf{\Gamma}}_{i,j})_N\}$ is asymptotically equivalent to the circulant matrix $\{(\mathbf{\Gamma}_{i,j})_N\}$ defined as

$$(\mathbf{\Gamma}_{i,j})_N \triangleq \mathbf{U}_N^\dagger (\mathbf{\Lambda}_{i,j})_N \mathbf{U}_N, \tag{8}$$

where $\mathbf{U}_N$ is the unitary discrete Fourier transform (DFT) matrix of dimension $N$ given by

$$\mathbf{U}_N(q,s) = \frac{1}{\sqrt{N}} e^{-\xi \frac{2\pi(q-1)(s-1)}{N}}, \quad q,s = 1, 2, \ldots, N, \tag{9}$$

and $(\mathbf{\Lambda}_{i,j})_N$ is a diagonal matrix with the $q$-th diagonal entry given by

$$(\mathbf{\Lambda}_{i,j})_N(q,q) = \Gamma_{i,j}\left(\frac{2\pi(q-1)}{N}\right), \quad q = 1, 2 \ldots, N. \tag{10}$$

Clearly, for a band-limited waveform with non-zero spectrum over its bandwidth, if the Nyquist sampling frequency $f_s = 2W$ is chosen, there is no deterministic zero in the spectrum of the sequence $\{\gamma_{i,j}(k), k \in \mathbb{Z}\}$ and the diagonal entries of $(\mathbf{\Lambda}_{i,j})_N$ are all non-zero bounded values. For large codeword length, we can approximate $(\hat{\mathbf{\Gamma}}_{i,j})_N$ by $(\mathbf{\Gamma}_{i,j})_N$. In the sequel, we may omit the superscript $(\cdot)_N$ for simplicity.

*Lemma 1:* If the shaping waveform has a sub-linear decaying rate in time, (i.e, $|\psi(t)| \leq a/|t/T_s|^\eta$, $a$ is a constant and $\eta > 1$), $\hat{\mathbf{\Gamma}}_{i,j}$ can be approximated by $\mathbf{\Gamma}_{i,j}$ with a negligible approximation error as $N \to \infty$.

The proof is given in Appendix B. Since the communication in this case is carried out over a strictly limited bandwidth $W$, according to the shift property of the DTFT, we get

$$\mathbf{\Lambda}_{i,j} = \mathbf{\Lambda}_0 \mathbf{E}(\hat{\tau}_{i,j}^{[i]}), \tag{11}$$

where $\hat{\tau}_{i,j}^{[i]} \triangleq \frac{\tau_{i,j}^{[i]}}{T_s}$, $\mathbf{\Lambda}_0$ is defined similar to $\mathbf{\Lambda}_{i,j}$ when $\hat{\tau}_{i,j}^{[i]} = 0$, and

$$\mathbf{E}(\hat{\tau}_{i,j}^{[i]}) = \texttt{diag}\left\{1, e^{-\xi \frac{2\pi}{N} \hat{\tau}_{i,j}^{[i]}}, e^{-\xi \frac{4\pi}{N} \hat{\tau}_{i,j}^{[i]}}, \ldots, e^{-\xi \frac{2(N-1)\pi}{N} \hat{\tau}_{i,j}^{[i]}}\right\}. \tag{12}$$

For sufficiently large values of N, by approximating $\hat{\mathbf{\Gamma}}_{i,j}$ and by $\mathbf{\Gamma}_{i,j}$ and substituting (11) to (8), we get

$$\underline{y}'_i = \mathbf{\Lambda}_0 \sum_{j=1}^{K} h_{i,j} \mathbf{E}(\hat{\tau}_{i,j}^{[i]}) \underline{x}'_j + \underline{z}'_i, \tag{13}$$

where $\underline{y}'_i, \underline{x}'_j, \underline{z}'_i$ are respectively the linear transformations of $\underline{y}_i, \underline{x}_j, \underline{z}_i$ by the DFT matrix, $\mathbf{U}$.

## B. When $\psi(t)$ Is a Time-Limited Waveform

Assume that $\psi(t)$ has a time support equal to $uT_s$, i.e., $\psi(t) = 0, \forall t \notin [0, uT_s]$. At the $j$-th transmitter, a codeword of length $N$, $\underline{x}_j = [x_j(0), x_j(1), \ldots, x_j(N-1)]^\intercal$, is supported by cyclic prefix and cyclic suffix symbols (CPS) each of length $u+1$ such that the first and the last $u+1$ symbols of $\underline{x}_j$ are respectively repeated at the end and at the beginning of this vector. The resulted vector, $\underline{x}_j^{cps} = [x_j(N-u-1), x_j(N-u), \ldots, x_j(N-1), x_j(0), x_j(1), \ldots, x_j(N-1), x_j(0), x_j(1), \ldots, x_j(u)]^\intercal$, of length $\ell = N + 2(u+1)$ is transmitted over the channel. The received signal at the $i$-th receiver node is given by

$$y_i(t) = \sum_{j=1}^{K} h_{i,j} \sum_{k=0}^{\ell-1} x_j^{cps}(k) \psi(t - kT_s - \tau_{i,j}) + z_i(t), \tag{14}$$

where $x_j^{cps}(k)$ is the $k$-th entry of $\underline{x}_j^{cps}$. This signal is passed through a filter matched to the desired link. The output of the matched filter sampled at $t = (k+1)T_s + \tau_{i,i}$, $k = 0, \ldots, \ell-1$, is given by

$$y_i(k) = \int_{kT_s + \tau_{i,i}}^{(k+u)T_s + \tau_{i,i}} y_i(t) \psi^*(t - kT_s - \tau_{i,i}) dt$$

$$= \sum_{j=1}^{K} h_{i,j} \sum_{q=-u}^{u} \gamma_{i,j}(q) x_j^{cps}(k+q) + z_i(k), \tag{15}$$

where $x_j^{cps}(q) = 0, \forall q < 0$, and

$$\gamma_{i,j}(q) = \int_0^{uT_s} \psi(t - qT_s + \tau_{i,j}^{[i]})\psi^*(t)dt, \tag{16}$$

$$z_i(k) = \int_{kT_s + \tau_{i,i}}^{(k+u)T_s + \tau_{i,i}} z_i(t)\psi^*(t - kT_s - \tau_{i,i})dt. \tag{17}$$

Each transmitted symbol of a stream is interfered by $(u-1)$ previous and $(u-1)$ future symbols (if not zero) of the same stream. It is also interfered by $2u-1$ symbols (if not zero) of each of the other transmitted streams. If the interfering stream is ahead of the desired stream, $u-1$ previous and $u$ future symbols of that stream interfere with the current symbol of the desired stream. However, if the interfering stream is behind the desired stream, $u$ previous and $u-1$ future symbols of that stream interfere with the current symbol of the desired stream. These can be verified by checking that $\gamma_{i,i}(u) = \gamma_{i,i}(-u) = 0$ and for $i \neq j$, $\gamma_{i,j}(u) = 0$ if $\tau_{i,j}^{[i]} < 0$, and $\gamma_{i,j}(-u) = 0$ if $\tau_{i,j}^{[i]} > 0$.

By discarding CPS symbols at the output of the matched filter, we obtain

$$\underline{y}_i = \sum_{j=1}^K h_{i,j}\mathbf{\Gamma}_{i,j}\underline{x}_j + \underline{z}_i, \tag{18}$$

where $\underline{x}_j = [x_j(0), x_j(1), \ldots, x_j(N-1)]^\mathsf{T}$, $\underline{y}_i = [y_i(u+1), y_i(u+2), \ldots, y_i(u+N)]^\mathsf{T}$, $\underline{z}_i = [z_i(u+1), z_i(u+2), \ldots, z_i(u+N)]^\mathsf{T}$, and $\mathbf{\Gamma}_{i,j}$ is the circulant convolution matrix of the generator sequence $\hat{\underline{\gamma}}_{i,j} = [\gamma_{i,j}(0), \gamma_{i,j}(1), \ldots, \gamma_{i,j}(u), 0, \ldots, 0, \gamma_{i,j}(-u), \ldots, \gamma_{i,j}(-1)]^\mathsf{T}$ of length $N$. Assuming $N \geq 2u$, $\mathbf{\Gamma}_{i,j}$ in general is given in equation (19) for all $i, j$. However, depending on the values of the relative asynchronous delays, $\gamma_{i,j}(-u)$ or $\gamma_{i,j}(u)$ might be zero.

$$\mathbf{\Gamma}_{i,j} = \begin{bmatrix} \gamma_{i,j}(0) & \cdots & \gamma_{i,j}(-u) & 0 & 0 & \ldots & 0 & \gamma_{i,j}(u) & \gamma_{i,j}(u-1) & \ldots & \gamma_{i,j}(1) \\ \gamma_{i,j}(1) & \cdots & \gamma_{i,j}(-u+1) & \gamma_{i,j}(-u) & 0 & \ldots & 0 & 0 & \gamma_{i,j}(u) & \ldots & \gamma_{i,j}(2) \\ \ddots & \ddots & \ddots & \ddots & \ddots & \ddots & \ddots & \ddots & \ddots & \ddots & \ddots \\ 0 & \cdots & 0 & \cdots & 0 & \gamma_{i,j}(u) & \cdots & \gamma_{i,j}(1) & \gamma_{i,j}(0) & \gamma_{i,j}(-1) \cdots & \gamma_{i,j}(-u) \\ \ddots & \ddots & \ddots & \ddots & \ddots & \ddots & \ddots & \ddots & \ddots & \ddots & \ddots \\ \gamma_{i,j}(-1) & \cdots & \gamma_{i,j}(-u) & 0 & \cdots & \cdots & 0 & \gamma_{i,j}(u) & \cdots & \gamma_{i,j}(1) & \gamma_{i,j}(0) \end{bmatrix}. \tag{19}$$

$\underline{z}_i$ is the colored noise vector at the $i$-th receiver with the covariance matrix $\mathbf{\Phi}_i = \sigma_i^2 \tilde{\mathbf{\Gamma}}_0$, where $\sigma_i^2$ is the variance of the samples of the additive white Gaussian noise at the $i$-th receiver and $\tilde{\mathbf{\Gamma}}_0$ is given in (20).

$$\tilde{\mathbf{\Gamma}}_0 = \begin{bmatrix} \gamma_{i,i}(0) & \gamma_{i,i}(-1) \cdots & \gamma_{i,i}(-u+1) & 0 & 0 & \ldots & 0 & \ldots & 0 \\ \gamma_{i,i}(1) & \cdots & \gamma_{i,i}(-u+2) & \gamma_{i,i}(-u+1) & 0 & \ldots & 0 & \ldots & 0 \\ \ddots & \ddots & \ddots & \ddots & \ddots & \ddots & \ddots & \ddots & \ddots \\ 0 & \cdots & 0 & 0 & \gamma_{i,i}(u-1) & \cdots & \gamma_{i,i}(0) & \cdots & \gamma_{i,i}(-u+1) \\ \ddots & \ddots & \ddots & \ddots & \ddots & \ddots & \ddots & \ddots & \ddots \\ 0 & 0 & \cdots & 0 & \cdots & 0 & \gamma_{i,i}(u-1) & \cdots & \gamma_{i,i}(0) \end{bmatrix}. \tag{20}$$

$\tilde{\mathbf{\Gamma}}_0$ is a Hermitian banded Toeplitz matrix of order $u$. This matrix is shown to be asymptotically equivalent to $\mathbf{\Gamma}_{i,i}$ given in (19) [19]. Since these matrices are Hermitian, their eigenvalues are all non-negative real numbers. Using properties of asymptotically equivalent Hermitian matrices in [19], it can be shown that the eigenvalues of $\tilde{\mathbf{\Gamma}}_0$ are all bounded[2].

---

[2]The correlated noise vector can be whitened by passing the received signal vector through a whitening filter. However, it is not necessary to do so here, because the bounded eigenvalues of $\tilde{\mathbf{\Gamma}}_i$ do not affect the total number of degrees of freedom of the underlying channel.

As can be seen from (18), due to the effect of the asynchronism among the users, the original quasi-static links with constant coefficients over a block as $h_{i,j}\mathbf{I}_N$, $i,j \in \{1,2,\ldots,K\}$, are converted to frequency-selective links with correlated coefficients and with gains over a block given by $h_{i,j}\mathbf{\Gamma}_{i,j}$.

*Remark 1:* Because both the previous and the future transmitted symbols of all the streams affect the current transmitted symbol of every single stream, it is necessary to add both the cyclic prefix and the cyclic suffix symbols to every transmitted frame in order to have circulant $\mathbf{\Gamma}_{i,j}$ matrices.

*Remark 2:* Since $\mathbf{\Gamma}_{i,j}$ is a circulant matrix, its eigenvalue decomposition is given by [19]

$$\mathbf{\Gamma}_{i,j} = \mathbf{U}^\dagger \mathbf{\Lambda}_{i,j} \mathbf{U}, \tag{21}$$

where $\mathbf{U}$ is the DFT matrix given in (9) and $\mathbf{\Lambda}_{i,j}$ is a diagonal matrix containing the elements of the DFT of the generator sequence of $\mathbf{\Gamma}_{i,j}$, i.e., $\mathbf{\Lambda}_{i,j} = \mathtt{diag}\{\lambda_{i,j}(0), \lambda_{i,j}(1), \ldots, \lambda_{i,j}(N-1)\}$, where

$$\lambda_{i,j}(k) = \sum_{q=0}^{N-1} \hat{\gamma}_{i,j}(q) e^{-\xi \frac{2\pi}{N} qk}, \quad k = 0\ldots, N-1, \tag{22}$$

and $\hat{\gamma}_{i,j}(q)$ is the $q$-th element of $\hat{\underline{\gamma}}_{i,j}$.

*Proposition 1:* For a waveform with non-zero spectrum over its bandwidth, $\mathbf{\Gamma}_{i,j}$ is a full rank matrix $\forall i,j \in \{1,2,\ldots,K\}$. In addition, its eigenvalues (the diagonal entries of $\mathbf{\Lambda}_{i,j}$) are bounded.

The proof is given in Appendix A. Matrix $\mathbf{\Lambda}_{i,j}$ contains the elements of the DFT of $\hat{\underline{\gamma}}_{i,j}$ or equivalently the samples of the DTFT of the sequence $\{\gamma_{i,j}(k), \forall k \in \mathbb{Z}\}$ given in (7) on its main diagonal. When $u \to \infty$ ($N \geq 2u$), $\gamma_{i,j}(k)$'s are the samples of a strictly limited bandwidth process of bandwidth $W = 1/2T_s$ (see the proof of Proposition 1). In this case, the shift property of the DTFT yields

$$\mathbf{\Lambda}_{i,j} = \mathbf{\Lambda}_0 \mathbf{E}(\hat{\tau}_{i,j}^{[i]}), \tag{23}$$

where $\hat{\tau}_{i,j}^{[i]} = \frac{\tau_{i,j}^{[i]}}{T_s}$, $\mathbf{\Lambda}_0$ is defined similar to $\mathbf{\Lambda}_{i,j}$ when $\hat{\tau}_{i,j}^{[i]} = 0$, and $\mathbf{E}(\hat{\tau}_{i,j}^{[i]})$ is given in (12).

*Lemma 2:* If the shaping waveform has a sub-linear decaying rate in time, for a finite value of $u$, the equality (23) still holds with a bounded approximation error which goes to zero as $u$ increases.

The proof is given in Appendix C. According to Lemma 2, for sufficiently large values of u, by decomposing the circulant matrices $\mathbf{\Gamma}_{i,j}$ on the DFT basis, we get

$$\begin{aligned} \underline{y}'_i &= \sum_{j=1}^{K} h_{i,j} \mathbf{\Lambda}_{i,j} \underline{x}'_j + \underline{z}'_i \\ &\simeq \mathbf{\Lambda}_0 \sum_{j=1}^{K} h_{i,j} \mathbf{E}(\hat{\tau}_{i,j}^{[i]}) \underline{x}'_j + \underline{z}'_i, \end{aligned} \tag{24}$$

where $\underline{y}'_i, \underline{x}'_j, \underline{z}'_i$ are respectively the linear transformations of $\underline{y}_i, \underline{x}_j, \underline{z}_i$ by the DFT matrix, $\mathbf{U}$.

## C. The Shaping Waveform

It was argued in the previous subsections that having a sub-linear decaying rate in time for the shaping waveform is a necessary and a sufficient condition to obtain the system models presented in (13) and (24) for band-limited and for time-limited waveforms, respectively. Hence, the raised-cosine waveform with a non-zero excess bandwidth, which has a decaying rate proportional to $1/|t/T_s|^3$ for large enough values of $t$, is a good candidate to be used in the structure of the proposed scheme (i.e., using the root-raised cosine waveform as the transmitter and the receiver filters). As can be seen, waveforms with faster decaying rate in time are more appealing to have a faster decaying approximation error with $u$.

*Remark 3:* According to the Paley-Wiener Theorem, the spectrum of a signal cannot be both time-limited and band-limited. Hence, using a truncated version of a waveform results in an unlimited support in the frequency domain [21]. For practical values of SNR ($\rho < \infty$), if a well-designed waveform with vanishing spectrums out of bandwidth $W$ is used, depending on the level of the noise at the receivers, one may choose $u$ sufficiently large such that the tails of the spectrum of the transmitted signals lie below the noise level. In this case, the system can be approximated with a bandlimited one. In theory, when $\rho \to \infty$, it is not possible to avoid bandwidth expansion when a finite support shaping waveform is used.

## III. PROOF OF THEOREM 1

### A. Proof of the First Part

Assume $\psi(t)$ with bandwidth $W = \frac{1}{2T_s}$ is used as the shaping waveform. Since the transmitted sequences are independent, the bandwidth of the received signal at each receiver node is also $W$. By taking the Fourier Transform of both sides of equation (3), we obtain

$$\begin{aligned}
Y_i(f) &= \int_{-\infty}^{\infty} \left( \sum_{j=1}^{K} h_{i,j} x_j(t - \tau_{i,j}) + z_i(t) \right) e^{-\xi 2\pi f t} dt \\
&= \sum_{j=1}^{K} h_{i,j} e^{-\xi 2\pi f \tau_{i,j}} \int_{-\infty}^{\infty} \sum_k x_j(k) \psi(t - kT_s) e^{-\xi 2\pi f t} dt + Z_i(f) \\
&= \sum_{j=1}^{K} h_{i,j} \Psi(f) e^{-\xi 2\pi f \tau_{i,j}} \sum_k x_j(k) e^{-\xi 2\pi f k T_s} + Z_i(f) \\
&= \sum_{j=1}^{K} h_{i,j} \Psi(f) e^{-\xi 2\pi f \tau_{i,j}} X_j(f) + Z_i(f) \\
&= \sum_{j=1}^{K} h'_{i,j}(f) X_j(f) + Z_i(f),
\end{aligned} \tag{25}$$

where $Z_i(f)$ is the Fourier Transform of $z_i(t)$, $h'_{i,j}(f) = h_{i,j}\Psi(f)e^{-\xi 2\pi f \tau_{i,j}}$, and $X_j(f)$ is the $2\pi$-periodic DTFT given in (7) of the transmitted sequence by the $j$-th transmitter. Equation (25) represents the mathematical model of a synchronous $K$-user interference channel with varying fading coefficients in the frequency domain. Since $\psi(t)$ has bandwidth $W$, $\Psi(f) = 0, \forall |f| > W$. For each specific value of $f \in [-W, W]$, the system is modeled as a constant synchronous interference channel which according to [12] has at most $K/2$ spatial DoF. Hence, the total number of complex DoF of the underlying channel is at most $W \times K/2$ per second which is tantamount to have at most $K/2$ spatial DoF per second per hertz. ∎

### B. Achieving Scheme, Asynchronous Interference Alignment

We propose an interference alignment algorithm which achieves the total $K/2$ DoF over the underlying quasi-static $K$-user asynchronous interference channel. Our scheme is similar to the vector interference alignment scheme proposed in [9] for time-varying channels wherein the fading coefficients of a link are independently chosen from a continuous probability distribution at the beginning of each symbol interval. In contrast, in our scenario, the communication links are quasi-static. As it was observed in previous sections, using the proposed signaling scheme, they are converted to ISI links and accordingly to time-varying channels. Hence, the proposed scheme provides the required channel variation for the interference alignment in quasi static scenarios. However, the channel coefficients of the links are not independently chosen from a distribution; rather, they are correlated in time. We show that even under channel correlation, the alignment task can almost surely be performed.

When $\psi(t)$ is a band-limited waveform, the signaling scheme proposed in Section II-A is used. In this case, the channel matrices, $\hat{\mathbf{\Gamma}}_{i,j}$'s, are approximated by their asymptotically equivalent matrices, $\mathbf{\Gamma}_{i,j}$'s, given in (8). However, when a time-limited waveform is used, the practical signaling scheme proposed in Section II-B is deployed. Having the same channel model with the same notation for both cases facilitates pursuing a single procedure for both scenarios. In the sequel, to avoid confusion, we detail the scheme for the case that a time-limited waveform is used and elaborate it for the other case when it is necessary.

Consider a scenario wherein the first user sends $(n+1)^\kappa$ streams of symbols to the first receiver via $(n+1)^\kappa$ distinct direction vectors each of length $N = (n+1)^\kappa + n^\kappa$, where $\kappa = (K-1)(K-2) - 1$. Each of the other transmitters sends $n^\kappa$ streams of symbols to the intended receiver via $n^\kappa$ distinct direction vectors of the same length $N$. The precoding matrix $\mathbf{V}_j$ at the $j$-th transmitter contains all the corresponding direction vectors as its columns. Each transmitted frame is supported by enough number of CPS symbols. The received signal vector at the $i$-th receiver node after discarding the CPS symbols is given by

$$\underline{y}_i = \sum_{j=1}^{K} h_{i,j} \mathbf{\Gamma}_{i,j} \mathbf{V}_j \underline{x}_j + \underline{z}_i. \tag{26}$$

To perform the alignment task, the precoding matrices should be designed such that the transmitted signals cast overlapping shadows at non-intended receivers while remaining distinct from the interference signals at the desired receivers. Thus, at the $i$-th receiver, the following condition should be satisfied.

$$\texttt{span}\ h_{1,2}\mathbf{\Gamma}_{1,2}\mathbf{V}_2 = \texttt{span}\ h_{1,3}\mathbf{\Gamma}_{1,3}\mathbf{V}_3 = \cdots = \texttt{span}\ h_{1,K}\mathbf{\Gamma}_{1,K}\mathbf{V}_K, \quad i=1,$$

$$\texttt{span}\ h_{i,j}\mathbf{\Gamma}_{i,j}\mathbf{V}_j \subset \texttt{span}\ h_{i,1}\mathbf{\Gamma}_{i,1}\mathbf{V}_1, \quad \forall\ i,j \neq 1,\ i \neq j,$$

where $\texttt{span}\ \mathbf{X}$ denotes the vector space spanned by matrix $\mathbf{X}$.

*Remark 4:* Since a vector space is closed under the scalar multiplication, it remains the same if the generator matrix is scaled by a constant value. Therefore, the fading coefficients do not play any role in aligning interference signals and can be neglected as long as they are non-zero.

For simplicity, we restrict ourself to the case that

$$\mathbf{\Gamma}_{1,2}\mathbf{V}_2 = \mathbf{\Gamma}_{1,3}\mathbf{V}_3 = \cdots = \mathbf{\Gamma}_{1,K}\mathbf{V}_K, \quad i=1, \tag{27}$$

$$\mathbf{\Gamma}_{i,1}^{-1}\mathbf{\Gamma}_{i,j}\mathbf{V}_j \prec \mathbf{V}_1, \quad \forall\ i,j \neq 1,\ i \neq j, \tag{28}$$

where $\prec$ indicates that the columns of the left hand side matrix are chosen from the columns of the right-hand side one. Equations (27) and (28) can respectively be simplified as follows.

$$\mathbf{V}_j = \mathbf{S}_j \mathbf{B}, \quad \forall j \in \{2,3,\ldots,K\}, \tag{29}$$

$$\mathbf{T}_{i,j}\mathbf{B} \prec \mathbf{A}, \quad \forall i,j \in \{2,3,\ldots,K\}, i \neq j, \tag{30}$$

where $\mathbf{A} = \mathbf{V}_1$, $\mathbf{B} = \mathbf{\Gamma}_{2,1}^{-1}\mathbf{\Gamma}_{2,3}\mathbf{V}_3$, $\mathbf{S}_j = \mathbf{\Gamma}_{1,j}^{-1}\mathbf{\Gamma}_{1,3}\mathbf{\Gamma}_{2,3}^{-1}\mathbf{\Gamma}_{2,1}$, $\forall j \in \{2,3,\ldots K\}$, and $\mathbf{T}_{i,j} = \mathbf{\Gamma}_{i,1}^{-1}\mathbf{\Gamma}_{i,j}\mathbf{S}_j$, $\forall i,j \in \{2,3,\ldots K\}, i \neq j$. Equation (30) characterizes $(K-1)(K-2)$ relations. To complete the alignment task, the union of the images of the precoding matrices at each receiver should span the whole vector space of dimension $N$, i.e., at the $i$-th receiver

$$\text{span }[\mathbf{V}_1, \mathbf{\Gamma}_{1,1}^{-1}\mathbf{\Gamma}_{1,3}\mathbf{V}_3] = \mathbb{C}^N, \quad i = 1, \tag{31}$$

$$\text{span }[\mathbf{V}_1, \mathbf{\Gamma}_{i,1}^{-1}\mathbf{\Gamma}_{i,i}\mathbf{V}_i] = \mathbb{C}^N, \quad i \neq 1. \tag{32}$$

In this case, each receiver can decode its desired data interference free by zer-forcing the interferences.

Let $\underline{w}$ be an arbitrary vector of length $N$ such that $\underline{w}' \triangleq \mathbf{U}\underline{w}$ contains only non-zero entries, where $\mathbf{U}$ is the DFT matrix of dimension $N$ given in (9). Let

$$\mathcal{V}_s \triangleq \left\{ \left( \prod_{i,j=2, i \neq j, (i,j) \neq (2,3)}^{K} \mathbf{T}_{i,j}^{\beta_{i,j}} \right) \underline{w}, \; \beta_{i,j} \in \{0, 1, \ldots, s\} \right\}. \tag{33}$$

$\mathbf{A}$ is chosen as a matrix which contains all vectors in $\mathcal{V}_n$. $\mathbf{B}$ is chosen as a matrix which contains all vectors in $\mathcal{V}_{n-1}$. Clearly, this choice of $\mathbf{A}$ and $\mathbf{B}$ satisfies (30). Since $\mathbf{B}$ is known, the precoding matrices, $\mathbf{V}_j, j = 2, 3, \ldots, K$, are obtained from (29).

*Proposition 2:* Matrix $\mathbf{V}_j$, $j = 1, 2, \ldots, K$, is a full column rank matrix almost surely. Moreover, the union of the images of the precoding matrices at each receiver node spans the whole vector space, $\mathbb{C}^N$, with probability one.

*Proof:* The proof is given for the first user. We show that the union of the images of the precoding matrices at the first receiver node spans the whole vector space. This a priori proves that $\mathbf{V}_1$ is a full column rank matrix. At receiver one, $[\mathbf{V}_1, \mathbf{\Gamma}_{1,1}^{-1}\mathbf{\Gamma}_{1,3}\mathbf{V}_3]$ and equivalently $[\mathbf{A}, \mathbf{FB}]$ should be full rank matrices, where $\mathbf{F} = \mathbf{\Gamma}_{1,1}^{-1}\mathbf{\Gamma}_{1,3}\mathbf{\Gamma}_{2,3}^{-1}\mathbf{\Gamma}_{2,1}$ which is simplified to $\mathbf{F} = \mathbf{U}^\dagger \mathbf{E}(\hat{\tau}_{1,3}^{[1]} + \hat{\tau}_{3,1}^{[2]})\mathbf{U}$ according to Remark 2 and Lemma 2. Let $\hat{\mathbf{F}} = \mathbf{E}(\hat{\tau}_f)$, where $\hat{\tau}_f = \hat{\tau}_{1,3}^{[1]} + \hat{\tau}_{3,1}^{[2]}$. $\hat{\mathbf{A}} \triangleq \mathbf{U}\mathbf{A}$ and $\hat{\mathbf{B}} \triangleq \mathbf{U}\mathbf{B}$, respectively, contain all vectors of $\hat{\mathcal{V}}_n$ and $\hat{\mathcal{V}}_{n-1}$, where $\hat{\mathcal{V}}_s$ is defined as follows,

$$\hat{\mathcal{V}}_s \triangleq \left\{ \mathbf{E}\left( \sum_{i,j=2, i \neq j, (i,j) \neq (2,3)}^{K} \beta_{i,j} \hat{\tau}_{t_{i,j}} \right) \underline{w}', \beta_{i,j} \in \{0, 1, \ldots, s\} \right\},$$

with $\hat{\tau}_{t_{i,j}} = \hat{\tau}_{1,j}^{[i]} + \hat{\tau}_{j,3}^{[1]} + \hat{\tau}_{3,1}^{[2]}$. Since the relative delays are independent and distinct continuous random variables, $\hat{\tau}_{t_{i,j}}$'s are independent and distinct random variables for all $i$ and $j$. Moreover, since the asynchronous delays are random variables of length less than a symbol interval, then $-3 < \hat{\tau}_{t_{i,j}} < 3, \forall i, j$. Note that $\hat{\tau}_f$ is independent of $\hat{\tau}_{t_{i,j}}$ for all $i, j \in \{2, 3, \ldots, K\}$. Similarly, $\hat{\tau}_f$ is a continuous random variable over $[-2, 2]$. It is seen that

$$[\mathbf{A}, \mathbf{FB}] = \mathbf{U}^\dagger[\hat{\mathbf{A}}, \hat{\mathbf{F}}\hat{\mathbf{B}}] = \mathbf{U}^\dagger \mathbf{W}\check{\mathbf{T}},$$

where $\mathbf{W}$ is a diagonal matrix with entries of $\underline{w}'$ on its main diagonal and $\check{\mathbf{T}} = [\tilde{\mathbf{A}}, \hat{\mathbf{F}}\tilde{\mathbf{B}}]$. $\tilde{\mathbf{A}}$ and $\tilde{\mathbf{B}}$ are defined similar to $\hat{\mathbf{A}}$ and $\hat{\mathbf{B}}$ when $\underline{w}'$ is a vector with all entries equal to one. Clearly, $[\mathbf{A}, \mathbf{FB}]$ and $\check{\mathbf{T}}$ have the same rank order. Let $\phi_{i,j} \triangleq e^{-\xi \frac{2\pi}{N} \hat{\tau}_{t_{i,j}}}$ and $\theta \triangleq e^{-\xi \frac{2\pi}{N} \hat{\tau}_f}$. For ease of understanding, we give matrix $\check{\mathbf{T}}$ for $K = 3$ in (34) in which $\phi = \phi_{3,2}$.

$$\check{\mathbf{T}}|_{K=3} = \begin{bmatrix} 1 & 1 & 1 & \cdots & 1 & 1 & 1 & 1 & \cdots & 1 \\ 1 & \phi & \phi^2 & \cdots & \phi^n & \theta & \theta\phi & \theta\phi^2 & \cdots & \theta\phi^{(n-1)} \\ 1 & \phi^2 & \phi^4 & \cdots & \phi^{2n} & \theta^2 & \theta^2\phi^2 & \theta^2\phi^4 & \cdots & \theta^2\phi^{2(n-1)} \\ \vdots & \vdots & \vdots & \cdots & \vdots & \vdots & \vdots & \vdots & \cdots & \vdots \\ 1 & \phi^{N-1} & \phi^{2(N-1)} & \cdots & \phi^{n(N-1)} & \theta^{N-1} & \theta^{N-1}\phi^{N-1} & \theta^{N-1}\phi^{2(N-1)} & \cdots & \theta^{N-1}\phi^{(N-1)(n-1)} \end{bmatrix}.$$

$$\tag{34}$$

In general, the $k$-th row of $\check{\mathbf{T}}$ has entries of the following forms.

$$\prod_{i,j=2, i\neq j, (i,j)\neq(2,3)}^{K} \phi_{i,j}^{(k-1)\beta_{i,j}}, \ \beta_{i,j} \in \{0, 1, \ldots, n\}, \ \ \text{or}$$

$$\left(\prod_{i,j=2, i\neq j, (i,j)\neq(2,3)}^{K} \phi_{i,j}^{(k-1)\beta_{i,j}}\right) \theta^{k-1}, \ \beta_{i,j} \in \{0, 1, \ldots, n-1\}.$$

One can see that $\check{\mathbf{T}}$ is a Vandermonde matrix of dimension $N$ with the generator vector $\check{\underline{t}}$ containing all the above entries for $k=2$. The determinant of $\check{\mathbf{T}}$ is given by the multiplication of all the elements of a set containing the difference of every two non-trivial entries of $\check{\underline{t}}$. Hence, if the entries of $\check{\underline{t}}$ are all distinct, $\check{\mathbf{T}}$ and accordingly $[\mathbf{A}, \mathbf{FB}]$ are full rank matrices. The entries of vector $\check{t}$ have the following form.

$$\theta^\alpha \left(\prod_{i,j=2, i\neq j, (i,j)\neq(2,3)}^{K} \phi_{i,j}^{\beta_{i,j}}\right), \ \beta_{i,j} \in \{0, 1, \ldots, s\}. \tag{35}$$

where $\alpha \in \{0, 1\}$ and $(\alpha, s) \in \{(0, n), (1, n-1)\}$. If two non-trivial elements of $\check{\underline{t}}$ are the same, we obtain

$$\theta^\alpha \prod_{i,j=2, i\neq j, (i,j)\neq(2,3)}^{K} \phi_{i,j}^{\beta_{i,j}} = \theta^{\alpha'} \prod_{i,j=2, i\neq j, (i,j)\neq(2,3)}^{K} \phi_{i,j}^{\beta'_{i,j}},$$

$$\Rightarrow \ (\alpha - \alpha')\hat{\tau}_f + \sum_{i,j=2, i\neq j, (i,j)\neq(2,3)} (\beta_{i,j} - \beta'_{i,j})\hat{\tau}_{t_{i,j}} = kN, \quad k \in \mathbb{Z}. \tag{36}$$

Note that each entry of $\check{\underline{t}}$ contains at least one parameter ($\theta$ or $\phi_{i,j}$) with different exponent ($\alpha$ or $\beta_{i,j}$) from that of the corresponding parameter in other entries. Therefore, the above equation does not trivially hold for $k = 0$. Moreover, since $\hat{\tau}_f$ and $\hat{\tau}_{t_{i,j}}$'s are continuous independent random variables respectively over $[-2, 2]$ and $[-3, 3]$ intevals, they do not satisfy equation (36) with probability one. Therefore, the determinant of $\check{\mathbf{T}}$ and accordingly the determinant of $[\mathbf{A}, \mathbf{FB}]$ are non-zero almost surely. ∎

*Corollary 1:* The proposed interference alignment scheme achieves the total DoF equal to $K/2$ over the constant K-user symbol-asynchronous interference channel almost surely.

*Proof:* According to Propositions 2, $(n+1)^\kappa + (K-1)n^\kappa$ independent information symbols are transmitted interference free over $(n+1)^\kappa + n^\kappa + 2(u+1)$ symbol intervals (the extra $2(u+1)$ symbol intervals are due to the transmission of CPS symbols). Hence, the efficiency factor of the transmission scheme is $\frac{(n+1)^\kappa + (K-1)n^\kappa}{(n+1)^\kappa + n^\kappa + 2(u+1)}$ regardless of the type of the shaping waveform used by transmitters. For a finite value of $u$, this factor becomes arbitrary close to $K/2$ for a large value of $n$. Hence, if a truncated version of the root-square raised cosine filters with a non-zero small excess bandwidth and sufficiently large time support is used, the total DoF arbitrary close to $K/2$ is achieved. Note that when a band-limited waveform is used, no CPS symbols are used and the efficiency factor is $\frac{(n+1)^\kappa + (K-1)n^\kappa}{(n+1)^\kappa + n^\kappa}$ which approaches to $K/2$ for a large $n$. In this case, the the root-square raised cosine filters with a non-zero small excess bandwidth provide the total DoF arbitrary close to $K/2$. ∎

*Remark 5:* It is obvious that the proposed scheme does not apply to the case when all the users are fully synchronous, i.e., $\tau_{i,j} = 0, \forall i, j$. One can check that inserting artificial delays at the transmitters or sampling the received signals with random delays at the receivers do not help and the scheme applies only to random independent delays. According to the definition of $\hat{\tau}_f$, if $\hat{\tau}_{1,3}^{[1]} = \hat{\tau}_{1,3}^{[2]}$, then $\hat{\tau}_f = 0$ and thus $\hat{\mathbf{F}} = \mathbf{E}(0) = \mathbf{I}_N$ resulting in $\check{\mathbf{T}}$ to be a rank deficient matrix. Clearly, when the users are synchronous, $\hat{\tau}_{1,3}^{[1]} = \hat{\tau}_{1,3}^{[2]}$ even if artificial delays are inserted at the transmitters or the received signals are sampled with random delays.

## IV. Asynchronous Interference Channel with Multiple Antenna Nodes

In a $K$-user interference channel with multiple antenna nodes, say $M$ antennas at each node, it is shown in [9] that the total DoF is upper bounded by $MK/2$. When the channel coefficients are time-varying, the underlying channel is converted to a single antenna nodes' interference channel with $MK$ independent users. In this case, the vector alignment scheme proposed in [9] is sufficient to achieve the total DoF of the channel. For the three-user constant interference channel this upper-bound is achieved by a vector alignment scheme proposed in [9] in a finite number of the channel uses. However, it is not known if in general for $K > 3$ the same scheme can achieve the total DoF of the channel under quasi-static assumption. For the constant channel scenario, the upper-bound is achievable using the real-interference-alignment technique proposed in [15] at infinite SNR and with infinite quantization's precision.

Similar to the single antenna nodes' interference channel, the asynchronism among the users can be deployed to perform the alignment task under the quasi-static assumption. In theory, the medium can be considered as an asynchronous interference channel with $MK$ independent single antenna users. In this case, by applying the same alignment scheme proposed in Section III-B, the total $MK/2$ DoF is achieved. However, since the antennas of each node are collocated, it is more practical to consider the same asynchronous delay for all links between each pair of transmitter-receiver nodes.

For such a scenario, assume that at the $j$-th transmitter node, $M$ independent streams of information symbols, $\underline{x}_j^{[p]}, p = 1, 2, \ldots, M$, each of length $s_j$, are independently precoded by a matrix $\mathbf{V}_j$ of size $N \times s_j$ and each of them is transmitted by one of the antennas. All the nodes use the same shaping waveform. Let $\mathbf{H}_{i,j}$ of dimension $M$ be the channel matrix between the $j$-th transmitter and the $i$-th receiver nodes. By deploying the proposed signaling scheme in sections II-A or II-B, the received signal model at the $q$-th antenna of the $i$-th receiver node is given by

$$\underline{y}_i^{[q]} = \sum_{j=1}^{K} \mathbf{\Gamma}_{i,j} \mathbf{V}_j \underline{X}_{i,j}^{[q]} + \underline{z}_i^{[q]}, q = 1, 2, \ldots, M, \tag{37}$$

where $\underline{X}_{i,j}^{[q]} = \sum_{p=1}^{M} h_{i,j}(q,p) \underline{x}_j^{[p]}$. $\underline{y}_i^{[q]}$ and $\underline{z}_i^{[q]}$ are respectively the received signal and the noise vectors at the output of the $q$-th antenna of the $i$-th receiver node. Matrix $\mathbf{\Gamma}_{i,j}$ represents the effect of the shaping waveform and the asynchronous delays between the $j$-th transmitter and the $i$-th receiver. $h_{i,j}(q,p)$ is the $(q,p)$-th entry of $\mathbf{H}_{i,j}$ (the entry at the $q$-th row and the $p$-th column). $\underline{X}_{i,j}^{[q]}$ is the image of the transmitted vectors from the $j$-th transmitter node at the $q$-th antenna of the $i$-th receiver node.

Since all links between collocated antennas experience the same asynchronous delay, matrix $\mathbf{\Gamma}_{i,j}$ is the same for all links between the $j$-th transmitter and the $i$-th receiver. Therefore, if the alignment task is performed at one of the received antennas, it is automatically performed at the other collocated antennas. Thus, the same precoding matrices as those designed in Section III-B for the single antenna nodes' scenario is sufficient to do the alignment task in the underlying multiple antenna nodes' scenario. Hence, $s_1 = (n+1)^\kappa$ and $s_j = n^\kappa$, $j = 2, 3, \ldots, K$, where $\kappa = (K-1)(K-2) - 1$. $N$ is chosen as $N = n^\kappa + (n+1)^\kappa$. In this case, each user can achieve the total $1/2$ DoF in time for large codeword length. By applying the zero-forcing filter at the output of each receive antenna, the interferences from other users are discarded. However, the transmitted signals from collocated antennas are aligned at each receiver antenna and they lie in the desired subspace for all collocated receiver antennas. The equivalent channel model in this case is the same as that of an $M \times M$ MIMO channel with total $M$ spatial DoF. Hence, each user can achieve the total $M/2$ DoF which is tantamount to achieving $MK/2$ DoF for the entire network. It is worth noting that, using the same alignment scheme proposed for the signal antenna nodes interference channel for the underlying multiple antenna nodes scenario causes the alignment task to be performed faster (i.e., in less number of channel uses) specially for large values of $M$.

## V. CONCLUSION

A symbol-asynchronous $K$-user interference channel with quasi-static fading coefficients was considered. It was argued that the total DoF of this channel is the same as that of the corresponding synchronous channel. We proposed an interference alignment scheme for the underlying constant interference channel by deploying the asynchronous delays among the received signals at each receiver node to achieve the total $MK/2$ DoF of this channel ($M$ is the number of antennas at each node) in the limit of the codeword's length. In the proposed scheme, there is no need to have the channel state information of the links at the transmitter side. Instead, the full state information of the asynchronous delays is required at all nodes. Although the asynchronous delays are assumed to be less than a symbol interval, the generalization to larger values of symbol-asynchronous delays is straightforward. If the maximum possible asynchronous delay among the users is less than $b$ symbol intervals, it is sufficient to support each transmitted frame by $u + b$ cyclic prefix and $u + b$ cyclic suffix symbols.

## APPENDIX A
## PROOF OF PROPOSITION 1

Define $\gamma(\tau) \triangleq \int_{-\infty}^{\infty} \psi(t-\tau)\psi^*(t)dt$ as the auto-correlation function of the shaping waveform $\psi(t)$. One can check that $\Gamma(f) = |\Psi(-f)|^2$, where $\Gamma(f), \Psi(f)$ are the Fourier Transforms of $\gamma(\tau), \psi(t)$, respectively. Therefore $\gamma(\tau)$ has the same bandwidth as that of the shaping waveform and it has also a non-zero spectrum over its bandwidth. Let $W_0$ be the frequency bandwidth of $\psi(t)$ when $u \to \infty$. For a finite value of $u$, let $\gamma^p(\tau) = \sum_{k=-\infty}^{\infty} \gamma(\tau + kT)$ be a periodic expansion of $\gamma(\tau)$ with period $T = NT_s$. We assume that $N \geq 2u$. Since the shaping waveform has a time support equal to $uT_s$, the auto correlation function has a time support smaller than or equal to $2uT_s$. $\hat{\underline{\gamma}}_{i,j} = [\gamma_{i,j}(0), \gamma_{i,j}(1), \ldots, \gamma_{i,j}(u), 0, \ldots, 0, \gamma_{i,j}(-u), \gamma_{i,j}(-u+1), \ldots, \gamma_{i,j}(-1)]$, which is the generator sequence of the matrix $\mathbf{\Gamma}_{i,j}$, represents the samples of $\gamma^p(\tau)$ in one period at $\tau = qT_s - \tau_{i,j}^{[i]}$, $q = 0, \ldots, N-1$, and with sampling frequency $f_s = 2W_0$. Since $\gamma(\tau)$ has a non-zero spectrum over its bandwidth, the DFT coefficients of the samples of $\gamma^p(\tau)$ over one period do not contain any deterministic zero. These coefficients appear as the diagonal entries of $\mathbf{\Lambda}_{i,j}$. Hence $\mathbf{\Lambda}_{i,j}$ and equivalently $\mathbf{\Gamma}_{i,j}$ are full rank matrices. In addition, since the sequence $\hat{\underline{\gamma}}_{i,j}$ is absolutely summable, the diagonal entries of $\mathbf{\Lambda}_{i,j}$ are bounded. ∎

## APPENDIX B
## PROOF OF LEMMA 1

Assume $\hat{\mathbf{\Gamma}}_{i,j}$ in (6) is approximated by $\mathbf{\Gamma}_{i,j}$ in (8). The approximation error matrix $\mathbf{\Upsilon}$ is defined as

$$\mathbf{\Upsilon} = \mathbf{\Lambda}_{i,j} - \mathbf{U}\hat{\mathbf{\Gamma}}_{i,j}\mathbf{U}^\dagger. \tag{38}$$

The $(q, s)$-th entries of $\mathbf{\Lambda}_{i,j}$ and $\mathbf{U}\hat{\mathbf{\Gamma}}_{i,j}\mathbf{U}^\dagger$ are respectively given by

$$\mathbf{\Lambda}_{i,j}(q, s) = \begin{cases} \sum_k \gamma_k e^{-\xi \frac{2\pi}{N}(q-1)k}, & q = s \\ 0, & q \neq s, \end{cases}$$

$$\mathbf{U}\hat{\mathbf{\Gamma}}_{i,j}\mathbf{U}^\dagger(q, s) = \frac{1}{N} \sum_{j=1}^{N} \sum_{i=1}^{N} \gamma_{i-j} e^{-\xi \frac{2\pi}{N}[(q-1)(i-1)-(s-1)(j-1)]}.$$

If $q = s$, we get

$$\mathbf{\Upsilon}(q, q) = \sum_k \gamma_k e^{-\xi \frac{2\pi}{N}(q-1)k} - \frac{1}{N} \sum_{j=1}^{N} \sum_{i=1}^{N} \gamma_{i-j} e^{-\xi \frac{2\pi}{N}(q-1)(i-j)}$$

$$= \sum_{k=-N+1}^{N-1} \frac{|k|}{N} \gamma_k e^{-\xi \frac{2\pi}{N} k(q-1)} + \sum_{|k| \geq N} \gamma_k e^{-\xi \frac{2\pi}{N} k(q-1)}.$$

Assuming that the shaping waveform is such that $\psi(t) \leq a/|t/T_s|^\eta$, where $a$ is a constant, we get

$$|\mathbf{\Upsilon}(q, q)| \leq \sum_{k=-N+1}^{N-1} \frac{|k|}{N} |\gamma_k| + \sum_{|k| \geq N} |\gamma_k|$$

$$\leq \frac{2a}{N} \sum_{k=1}^{N-1} \frac{1}{k^{\eta-1}} + 2a \sum_{k=N}^{\infty} \frac{1}{k^\eta} \quad (39)$$

One can check that for $\eta > 1$, $\lim_{N \to \infty} \frac{1}{N} \sum_{k=1}^{N} \frac{1}{k^{\eta-1}} = 0$, because for all $N$ and $\eta > 0$,

$$0 = \lim_{N \to \infty} \frac{1}{N} \int_1^N \frac{1}{x^\eta} dx < \lim_{N \to \infty} \frac{1}{N} \sum_{k=1}^{N} \frac{1}{k^\eta} < \lim_{N \to \infty} \frac{1}{N} \int_1^N \frac{1}{(x-1)^\eta} dx = 0.$$

Hence, the first term on the right hand side of (39) goes to zero for large values of $N$. It is also known that the $\eta$-series $\sum_{k=1}^{\infty} 1/k^\eta$ is convergent for all $\eta > 1$. Hence, the second term is also bounded for $\eta > 1$ and vanishes to zero for large values of $N$. Therefore, for large codeword length, $|\mathbf{\Upsilon}(q, q)|$ is bounded and vanishes to zero if $\eta > 1$. When $q \neq s$, we get

$$\mathbf{\Upsilon}(q, s) = -\mathbf{U}\hat{\mathbf{\Gamma}}_{i,j}\mathbf{U}^\dagger(q, s)$$

$$= -\frac{1}{N} \sum_{j=1}^{N} \sum_{i=1}^{N} \gamma_{i-j} e^{-\xi \frac{2\pi}{N}[(q-1)(i-1)-(s-1)(j-1)]}$$

$$= -\frac{1}{N} \sum_{k=1}^{N-1} \gamma_k e^{-\xi \frac{2\pi}{N}(q-1)k} \sum_{j=k}^{N-1} e^{-\xi \frac{2\pi}{N}(q-s)(j-k)}$$

$$- \frac{1}{N} \sum_{k=-N+1}^{-1} \gamma_k e^{\xi \frac{2\pi}{N}(s-1)k} \sum_{j=-k}^{N-1} e^{-\xi \frac{2\pi}{N}(q-s)(j+k)}$$

$$= \frac{1}{N} \sum_{k=1}^{N-1} \gamma_k e^{-\xi \frac{2\pi}{N}(s-1)k} \sum_{j=N-k}^{N-1} e^{-\xi \frac{2\pi}{N}(q-s)j}$$

$$+ \frac{1}{N} \sum_{k=-N+1}^{-1} \gamma_k e^{-\xi \frac{2\pi}{N}(q-1)k} \sum_{j=N+k}^{N-1} e^{-\xi \frac{2\pi}{N}(q-s)j}.$$

Assuming that the shaping waveform is such that $\psi(t) \leq a/|t/T_s|^\eta$, where $a$ is a constant, then we get

$$|\Upsilon(q,s)| \leq \frac{1}{N}\sum_{k=-N+1}^{-1}|k||\gamma_k| + \frac{1}{N}\sum_{k=1}^{N-1}|k||\gamma_k|$$

$$\leq \frac{2a}{N}\sum_{k=1}^{N-1}\frac{1}{k^{\eta-1}}$$

Similarly for $\eta > 1$, $|\Upsilon(q,s)|$ is bounded and vanishes to zero for large values of $N$. ∎

## APPENDIX C
### PROOF OF LEMMA 2

Let $x(t)$ be a signal with bandwidth $W$. We assume that $x(t)$ is a decaying function of $|t/T_s|$ such that $|x(t)| \leq \frac{a}{|t/T_s|^\eta}$ for large enough values of $t$ and $\eta > 0$. Let $\{\hat{x}(k), k \in \mathbb{Z}\}$ be the sequence of samples of this signal at $t = kT_s$, where $T_s \leq \frac{1}{2W}$. Let $\{x(k), k \in \mathbb{Z}\}$ be the sequence of samples of $x(t)$ at $t = kT_s - \tau$, $0 < \tau < T_s$. According to the shift property of the DTFT, we get

$$X(\omega) = \hat{X}(\omega)e^{-\xi\omega\tau}, \tag{40}$$

where $\hat{X}(\omega) = \sum_k \hat{x}(k)e^{-\xi\omega k}$ and $X(\omega) = \sum_k x(k)e^{-\xi\omega k}$ are the DTFT of the sequences $\{\hat{x}(k), k \in \mathbb{Z}\}$ and $\{x(k), k \in \mathbb{Z}\}$, respectively. Define $A_k \triangleq x(k)e^{-\xi\omega k}$, $B_k \triangleq \hat{x}(k)e^{-\xi\omega(k+\tau)}$, and $\epsilon = \sum_{k=-\infty}^{-u-1}(B_k - A_k) + \sum_{k=u+1}^{\infty}(B_k - A_k)$. From (40), we get

$$\sum_{k=-u}^{u} A_k = \sum_{k=-u}^{u} B_k + \epsilon$$

$|\epsilon|$ is upper bounded as follows.

$$|\epsilon| = |\sum_{k=-\infty}^{-u-1}(B_k - A_k) + \sum_{k=u+1}^{\infty}(B_k - A_k)|$$

$$\leq \sum_{k=-\infty}^{-u-1}(|B_k| + |A_k|) + \sum_{k=u+1}^{\infty}(|B_k| + |A_k|)$$

$$= \sum_{k=-\infty}^{-u-1}(|\hat{x}_k| + |x_k|) + \sum_{k=u+1}^{\infty}(|\hat{x}| + |x_k|)$$

$$\leq \sum_{k=u+1}^{\infty}\frac{2a}{k^\eta} + \frac{a}{(k-\hat{\tau})^\eta} + \frac{a}{(k+\hat{\tau})^\eta}$$

$$\leq 4a\sum_{k=u}^{\infty}\frac{1}{k^\eta},$$

where $\hat{\tau} = \tau/T_s$. One can see that if $\eta > 1$, $|\epsilon|$ is bounded and vanishes to zero when as $u$ increases. ∎